\documentclass[sigconf]{acmart}

\AtBeginDocument{%
  \providecommand\BibTeX{{%
    \normalfont B\kern-0.5em{\scshape i\kern-0.25em b}\kern-0.8em\TeX}}}

\setcopyright{acmlicensed}
\copyrightyear{2025}
\acmYear{2025}
\acmDOI{XXXXXXX.XXXXXXX}

\usepackage{amsmath}
\usepackage{hhline}
\usepackage{booktabs}
\usepackage{tabularx}
\usepackage{graphicx}
\usepackage{siunitx}
\usepackage{multirow}
\usepackage{tikz}
\usetikzlibrary{tikzmark,calc}
\newcommand{\sys}{\text{E-QUARTIC}}

\acmConference[ASP-DAC 2025]{ACM Asia South Pacific Design
Automation Conference}{Jan. 20--23, 2025}{Tokyo, Japan}
\acmISBN{978-1-4503-XXXX-X/18/06}

\begin{document}

\title{E-QUARTIC: \underline{E}nergy \underline{E}fficient \underline{E}dge \underline{E}nsemble of Convolutional Neural Networks for Resource-Optimized Learning}


\author{Le Zhang, Onat Gungor, Flavio Ponzina, Tajana Rosing}
\affiliation{
  \institution{University of California San Diego}
  \city{La Jolla}
  \state{CA}
  \country{USA}}
\email{{lez014, ogungor, fponzina, tajana}@ucsd.edu}


\begin{abstract}
Ensemble learning is a meta-learning approach that combines the predictions of multiple learners, demonstrating improved accuracy and robustness. Nevertheless, ensembling models like Convolutional Neural Networks (CNNs) result in high memory and computing overhead, preventing their deployment in embedded systems. These devices are usually equipped with small batteries that provide power supply and might include energy-harvesting modules that extract energy from the environment. In this work, we propose \emph{\textbf{E}-QUARTIC}, a novel \textbf{E}nergy \textbf{E}fficient \textbf{E}dge \textbf{E}nsembling framework to build ensembles of CNNs targeting Artificial Intelligence (AI)-based embedded systems. Our design outperforms single-instance CNN baselines and state-of-the-art edge AI solutions, improving accuracy and adapting to varying energy conditions while maintaining similar memory requirements. Then, we leverage the multi-CNN structure of the designed ensemble to implement an energy-aware model selection policy in energy-harvesting AI systems. We show that our solution outperforms the state-of-the-art by reducing system failure rate by up to 40\% while ensuring higher average output qualities. Ultimately, we show that the proposed design enables concurrent on-device training and high-quality inference execution at the edge, limiting the performance and energy overheads to less than 0.04\%.
\end{abstract}




\maketitle

\section{Introduction}

Energy-harvesting edge machine learning (ML) systems aim to deliver real-time and reliable inference results without human intervention~\cite{gobieski2019intelligence}. These systems can adapt to new data distributions from sensor inputs through on-device retraining~\cite{lee2019intermittent}. Due to the practical application potentials, these systems are ideal for deployment in remote or inaccessible locations for tasks such as 
wildlife conservation~\cite{li2016wildlife}, smart cities technologies~\cite{akan2017smartcity}, and wearable devices~\cite{chong2019wearable}. However, implementing high-accuracy but resource-intensive ML models like convolutional neural networks (CNNs) poses challenges due to limited computational and memory resources~\cite{teerapittayanon2016branchynet}. Recent research addresses these challenges through two main aspects: optimizing ML models for resource efficiency and enhancing energy management~\cite{islam2020zygarde,lee2019intermittent,park2023energy,jeon2023harvnet, luo2023efficient}.

\begin{figure}[tp]
    \centering
    \includegraphics[width=\linewidth]{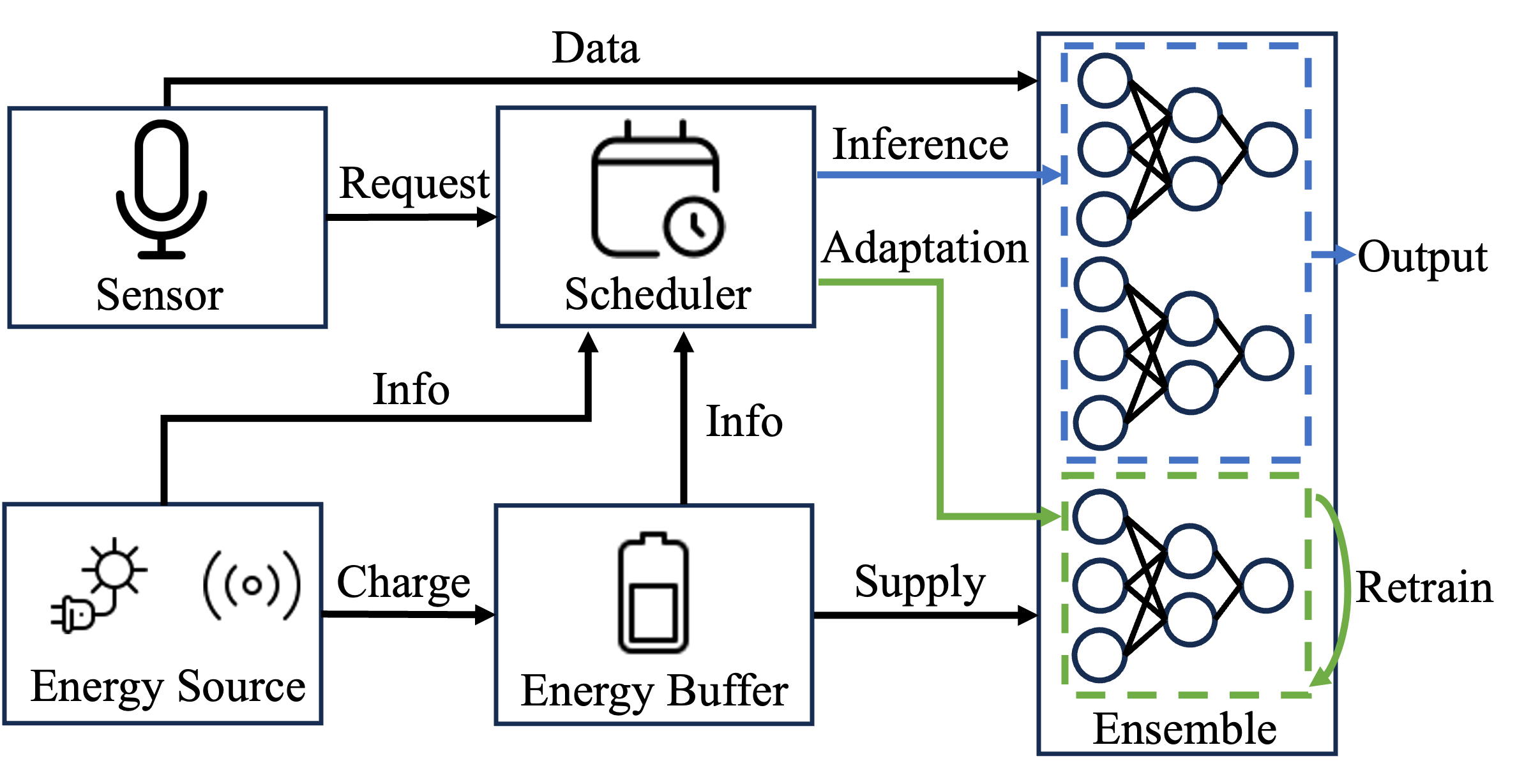}
    \vspace{-0.8cm}
    \caption{\sys~Overview}
    \label{fig:overview}
    \vspace{-0.5cm}
\end{figure}

Several methods for energy-optimized model design 
have been proposed~\cite{islam2020zygarde, park2023energy, jeon2023harvnet}. Lightweight neural networks~\cite{islam2020zygarde} facilitate edge AI, but often result in low accuracy. Selecting CNN instances from a model pool~\cite{park2023energy} and early-exit neural networks~\cite{jeon2023harvnet} can instead achieve high accuracy, but introduce memory and computational overhead that may prevent their deployment in tinyML systems. Moreover, their architectural design makes them inflexible for on-device training. These issues highlight the need to balance the adaptability of small neural networks with the capability of large models. Ensemble learning is a well-known approach for improving accuracy and robustness which combines the outputs of various "weak learners". However, ensemble learning presents challenges in resource-constrained embedded systems, e.g., increased memory demands, and higher energy consumption. To address these concerns, previous research efforts~\cite{wen2020batchensemble, ponzina2021e2cnn} explore resource-optimized ensemble algorithms by balancing the high performance of ensemble models and the practical constraints of limited resources. 

Apart from memory-optimized model design, recent works also enhance energy management and scheduling policies, aiming to optimize the runtime and efficiency performance and of these models~\cite{islam2020zygarde,lee2019intermittent,park2023energy,jeon2023harvnet}. Various dynamic scheduling algorithms have been developed to dynamically assign inference and training tasks based on system-level information such as remaining battery charge and harvested energy. These algorithms range from short-term priority-based scheduling~\cite{islam2020zygarde} to long-term planning with reinforcement learning agents~\cite{lee2019intermittent,park2023energy,jeon2023harvnet}. However, these approaches often face challenges: either lacking in long-term forecasting accuracy or incurring significant memory and computational costs. Ideally, a task scheduling agent on embedded systems should consider both performance and resource efficiency: a complex scheduler may offer better performance but is prone to create unacceptable runtime, energy, and memory overhead due to the limited resources of ultra-low-power systems. Balancing the decision-making performance and resource overhead is crucial in designing a resource-friendly scheduler.

In this paper, we propose ~\sys, a novel \underline{E}nergy \underline{E}fficient \underline{E}dge \underline{E}nsembling framework targeting energy-harvesting embedded systems. First, E-QUARTIC introduces a novel design methodology to construct energy-adaptive ensembles of CNNs that do not increase baseline memory and computing requirements. This is achieved by combining filter-level pruning methods with boosting algorithms for training the weak learners composing E-QUARTIC. Second, we leverage this novel ensemble-based design to propose an energy-aware scheduler that tailors inference executions based on energy considerations, ultimately improving the trade-off between output quality and energy consumption. We implement our framework on an energy-harvesting microcontroller for on-device dynamic inference and adaptation and evaluate E-QUARTIC in energy-constrained settings.
Figure~\ref{fig:overview} illustrates an overview of E-QUARTIC. Energy-adaptive ensembles of CNNs are first generated offline and implemented on microcontrollers. During runtime on-device execution, for each input sample, the energy-aware scheduler dynamically assigns inference and retraining stages considering a specific number of weak learners to generate the ensemble inference output or to retrain themselves. These decisions are made dynamically based on battery level and harvested power level. 
We summarize our contributions as follows:
\begin{itemize}
    \item We propose E-QUARTIC, a novel ensembling method to build a high-accuracy \textit{boosting-based energy-adaptive} ensemble of CNNs. E-QUARTIC combines pruning and ensemble learning to generate ensembles that do not increase computation complexity and reduce memory requirements compared to single-instance CNN baselines. 
    \item We enrich E-QUARTIC with an \textit{energy-aware Q-learning-based scheduler} to dynamically select the number of weak learners during inference to improve the trade-off between model accuracy and energy efficiency. 
    \item We perform on-device evaluations on an indoor energy harvesting dataset for E-QUARTIC against the state-of-the-art methods. E-QUARTIC achieves output qualities comparable to or slightly higher than state-of-the-art approaches while reducing system failure rate by up to 40\%. 
\end{itemize}




\section{Related Works}

\textbf{Ensemble Learning:} Ensemble learning enhances model accuracy and stability by combining multiple weak learners~\cite{dietterich2000ensemble}. 
However, it is often considered a resource- and compute-intense approach, thus limiting its real-life deployment in resource-constrained embedded systems. 
Recent research focuses on ensembling neural networks. The authors of~\cite{ponzina2021e2cnn} use random filter pruning~\cite{li2016pruning} to reduce the number of filters in a baseline single-instance CNN model, ultimately obtaining smaller weak learners that together meet memory constraints and improve model robustness against memory faults for low-power embedded systems. This method only guarantees memory equivalence (instead of energy/performance equivalence) and the use of random pruning allows operating on untrained CNN models but may result in unbalanced CNN architectures. 
An adaptation of the AdaBoost algorithm including transfer learning to reduce computational costs in training is instead proposed in~\cite{taherkhani2020adaboostcnn}. 
Yet, these methods reveal certain shortcomings. First, most ensemble methods are impractical for embedded systems, because of their higher resource requirements that prevent their deployment and evaluation on low-power embedded systems like energy-harvesting devices deployed in dynamic environments. Second, the design of hardware-aware ensembles of CNNs as the one proposed in~\cite{ponzina2021e2cnn} does not include state-of-the-art ensemble learning algorithms like ensemble pruning, but only rely on random pruning approaches that may result in suboptimal output quality.

\noindent
\textbf{Energy-Efficient Machine Learning:} To satisfy the limited battery capacity in edge computing environments, embedded ML systems are designed to balance model complexity and energy consumption. For instance, previous works apply early-exit neural networks with neural architecture search (NAS) to optimize the trade-off between accuracy and energy consumption~\cite{odema2021eexnas, jeon2023harvnet}. However, early exits are essentially built upon a single, large neural network that can hardly be leveraged to improve efficiency in on-device training. Another study~\cite{park2023energy} selects a neural network model from a pool of models with diverse energy budgets. However, their approach leads to a significant increase in memory usage due to storing multiple neural networks. This scaling issue, where memory usage grows with the granularity of energy management, poses a limitation for microcontrollers. 

\noindent
\textbf{Harvested Energy Management:} Energy-harvesting embedded ML systems are equipped with energy harvesters to perform on-device inference and training tasks by utilizing energy extracted from the environments~\cite{islam2020zygarde,lee2019intermittent}. Frequent energy depletion will lead to the failure of systems, impairing their ability to sense ambient events and deliver reliable inference results. Consequently, energy management policies assume pivotal roles in this context. They can be divided into two categories: Power-Neutral Operations (PNO)~\cite{islam2020zygarde} and Energy-Neutral Operations (ENO)~\cite{jeon2023harvnet}. 
PNO devices store minimal energy for immediate use, while ENO devices, with larger energy storage, ensure stable and prolonged performance, ideal for reliable predictions. We focus on the prevalent cases of ENO, utilizing buffered energy in super-capacitors to ensure reliable inference and adaptive applications via scheduling. 
Recent research applies diverse decision-making algorithms to develop task schedulers for ENO devices. 
In~\cite{park2023energy}, the authors leverage deep reinforcement learning agents and policy neural networks to select models. However, executing policy neural networks on embedded systems requires substantial amounts of time, memory, and energy. Q-learning is a model-free reinforcement learning approach where an agent learns to select optimal actions in various states based on future reward predictions~\cite{watkins1992q}. In~\cite{jeon2023harvnet}, authors use Q-learning agents to select early exit points in neural networks, leading to improved robustness and battery life.  
While Q-learning is efficient and computationally light, integrating it with long-term state information limits its suitability for short-interval, high-frequency inference tasks.

\noindent
\textbf{On-device Training:} 
On-device training is deemed a critical feature for edge machine learning systems to ensure consistent and reliable inference outcomes. However, the complexity and computational demand of online retraining pose significant challenges, especially on energy-limited systems where the scale of neural network models makes continuous updating impractical. A layer-wise retraining approach specifically for PNO applications, which schedules the retraining of the last fully-connected layers is introduced in~\cite{lee2019intermittent}. This strategy leverages dynamic energy planning to optimize the limited harvested energy for necessary adaptations. Nonetheless, this solution doesn't fully address the issue of scalability, as more sophisticated neural networks typically have larger and more complex layer structures. Additionally, the energy consumed by on-device training can strain the available energy storage and potentially compromise the system's inference reliability. In contrast, our design counters scalability issues by utilizing simpler weak learners instead of a large, complex neural network, significantly reducing energy demands for retraining. Moreover, a concurrent inference/training strategy is employed to enable on-device training while ensuring reliable and accurate inference results.

\section{E-QUARTIC Generation} \label{sec:ensemble-generation}
We provide an overview of E-QUARTIC design in Figure~\ref{fig:overview}. During the offline design phase, E-QUARTIC uses a baseline single-instance CNN architecture to build a \textit{boosting-based energy-adaptive ensembles of CNNs} showing (i) high accuracy, (ii) no computational complexity overhead, and (iii) lower memory requirements. 
In this section, we describe how the E-QUARTIC ensemble-based structure is derived.

\noindent
\textbf{Pruning:}
The computational complexity of CNNs is mainly tied to the number of filters in the convolutional layers. This observation suggests a novel approach to constructing an ensemble of CNN learners that can fit into computational budgets and constrained memory. The computational complexity can be estimated using multiply-accumulate operations (MACs).
To derive an ensemble of $N$ CNNs, we employ a pruning-replication approach similar to the one presented in~\cite{ponzina2021e2cnn}, which suggests pruning the target single-instance model by a factor $N$ so that the generated ensemble containing $N$ instance will not incur computational overheads. In our experiments, we observe the best results across different benchmarks when setting $2\leq N \leq5$. However, in contrast to~\cite{ponzina2021e2cnn} where random pruning is used to reduce only memory requirements, we also focus on reducing \emph{computational complexity (MACs)}, targeting for more adaptive \emph{runtime energy consumption}. Besides, we employ a better iterative L2-norm filter-wise pruning method~\cite{li2016pruning} to prune the initial single-instance model to obtain a pruned CNN with $\frac{1}{N}$ of the original number of MACs. 
Since the pruning of a single filter significantly reduces the output size, it also reduces the size of filters in the next layer because they will be then applied to lower-dimensional inputs. Therefore, our pruning method not only meets the constraint on MACs but also results in significant memory savings which will be discussed in Section~5.
To effectively achieve and leverage diversity, we first define a candidate pool of weak learners of size $M>N$.
Each of them is generated starting from a target baseline CNN model that is randomly initialized and trained. The model is then pruned and retrained for multiple iterations by gradually reducing the number of filters until it achieves the desired $\frac{1}{N}$ MACs. This process results in a candidate pool of $N$ weak learners, each of them having $\frac{1}{N}$ MACs and $< \frac{1}{N}$ memory requirements of the single-instance baseline model.

\begin{figure}[tp]
    \centering
    \includegraphics[width=\linewidth]{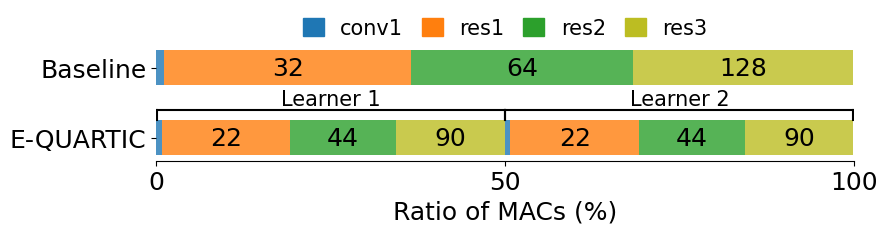}
    \caption{Ratio of multiply-accumulate operations (MACs) for each layer of ResNet-8 (Baseline) and an~\sys~design including two CNN instances.}
    \label{fig:macs-comparison}
    \vspace{-0.4cm}
\end{figure}

As an example, we illustrate in Figure~\ref{fig:macs-comparison} the MACs patterns of ResNet-8~\cite{he2016resnet} and that of a derived ensemble of two learners. The four colors represent the computational complexity of the initial convolutional layer (blue) and three residual blocks (orange, dark green, and light green), with numbers indicating the number of filters. The figure highlights how we create an ensemble (of two learners, in this example) that does not result in computational overhead when compared to the baseline CNN implementation.
We refer to the generated~\sys~configuration as a 2-instance ensemble, an ensemble of CNNs of size two (i.e., composed of two instances).


\noindent
\textbf{Boosting:} 
To enhance the overall accuracy of the ensemble by promoting diversity, we enrich the previously described ensemble design with a boosting method to enhance ensemble accuracy by prioritizing learning from previously misclassified samples~\cite{taherkhani2020adaboostcnn}. 
TO this end, we use AdaBoost~\cite{taherkhani2020adaboostcnn} to emphasize learning from the misclassified samples of the previously deployed learners. 
This is achieved by adjusting the loss contribution for each training sample through the assignment of individual "sample weight".
Initially, each training data $(x_i,y_i)$ is assigned a sample weight parameter $w_i^m$ with $m$ representing the number of candidate weak learners in the pool. At the beginning of training, when no weak learners are present, all $w_i^m$ are set to 1, such that $w_i^0=1$. 
As each weak learner completes its training and joins the pool, sample weights are updated according to the equation:
\begin{equation}
    w_i^m=w_i^{m-1}exp(-\alpha y_i^T log(f^m(x_i)))
    \label{eq:weight-update}
\end{equation}
where $w_i^{m-1}$ is the previous sample weight, $\alpha$ is the learning rate, and $f^m$ is the most recent weak learner added in the pool. Following the update, the weights are normalized to follow a Gaussian distribution. Following Equation~\ref{eq:weight-update}, sample weight values increase after misclassifications and decrease after correct prediction. This adjustment ensures the previous misclassified samples have larger weighted loss values, and the subsequent learner can effectively learn from these samples. The subsequent weak learner is then initialized, trained, and iteratively pruned using the updated sample weights. This process is repeated until $M$ weak learners are selected.

\noindent
\textbf{Backfitting:} 
The problem of refining a pool of $M$ weak learners to an ensemble of $N$ weak learners is formally defined as ensemble pruning~\cite{margineantu1997pruning}. 
Specifically, to ensure high accuracy in the resulting ensemble, backfitting~\cite{margineantu1997pruning} is employed as a heuristic solution to approximate the brute-force search of the optimal subset of $N$ weak learners. This procedure starts with greedily including learners that maximize subset accuracy. Besides the greedy selection process, previously selected learners are re-evaluated by comparing them with candidates in the pool. Learners found to be less effective are replaced with those contributing to higher accuracy. 
This process terminates once the ensemble includes exactly $N$ weak learners, making it equivalent to the original single-instance CNN from computing and energy perspectives.
Finally, we define each learner's weight as $a_m = \frac{1}{2}log(\frac{1-e_m}{e_m})$, where $e_m$ represents the error rate of the learner.
A weighted voting method is employed to derive the final prediction: $\hat{y}=\sum_{m=1}^{N} a_m f^m(x)$.

\section{E-QUARTIC Adaptation to Dynamic Energy Environments} \label{sec:inference}

Focusing on online executions, we introduce an energy-aware scheduler trained using simulated energy traces. At runtime, energy harvested from ambient sources powers the system. For each collected sample, the proposed E-QUARTIC scheduler dynamically allocates inference and retraining tasks to individual weak learners, considering relevant system information like battery levels and harvested energy. We describe how we leverage the described \sys~design in energy-harvesting dynamic environments for efficient inference by designing an \emph{energy-aware Q-learning-based scheduler}. Then, we demonstrate how to utilize it for energy-efficient concurrent inference and training for on-device adaptation.

\noindent
\textbf{Energy-Aware Inference:}
We develop an \emph{energy-aware Q-learning-based} to run inference tasks based on battery levels and harvested energy (\emph{Scheduler} in Figure~\ref{fig:overview}). The key observation is that we can exploit the ensemble-based architecture of E-QUARTIC by running only a subset of the deployed $N$ learners to save energy. Therefore, for each inference execution, the proposed scheduler determines the number $k<N$ of weak learners to run based on system energy information, thus trading off accuracy for efficiency.
To maximize output quality for any subset $k$ of the $N$ deployed learners, learners are sorted by descending accuracy\footnote{Accuracy measured during the training phase on the evaluation set.} to set their execution sequence.

To properly select $k$, the proposed scheduler uses a Q-learning approach based on the reward function shown in Equation~\ref{eq:rewards} and the state space summarized in Table~\ref{tab:state-space}.
$E_{now}$ and $E_{last}$ represent the current battery level and the mean battery level in the last 10 inference executions in four discrete states: depleted (less than the amount for executing one weak learner), low (less than 1/2 of the max capacity), high (greater than 1/2 of the max capacity), and full (equal to the max capacity). $P_{harv}$ represents the current harvesting power amount in three discrete states: low, mid, and high. $L$ keeps track of the number of already executed weak learners in the current inference, while $R$ is a binary indicator of an active inference request. The action space of our scheduler is composed of just two possible actions: executing the next learner, if there exists a next one ($L<N$), or not. When the selected action is to execute the next learner (i.e., $a=1$), the system schedules a new learner for execution. In the opposite case (i.e., $a=0$), the system aggregates the outputs of the already executed learners to produce the E-QUARTIC prediction and then moves to a sleeping mode to reduce energy consumption, waiting for the next inference request. The reward function is defined in Equation~\ref{eq:rewards} and determines the action to be taken (i.e., execute or not a new learner). $\Delta_{acc}$ represents the expected accuracy improvement achieved by executing the next learner ($a=1$). This improvement is calculated based on the evaluation dataset during the design phase. An energy term, parameterized by $\beta$, is applied to penalize energy depletion due to an additional learner execution, with a larger penalty incurred as the current battery level $E_{now}$ falls well below the maximum capacity $E_{max}$. Additionally, a penalty term $p_{miss}$ is introduced to discourage missing inference tasks. This occurs when the system fails to fulfill an inference request, meaning an inference request exists ($R=1$), but the system does not execute it ($a=0$). 

The proposed scheduling strategy stands out for its lightweight computing requirements and its fine-grained energy management that reassesses the battery level and power state following each learner's execution. By doing so, it allows for immediate and dynamic adjustments to energy consumption and operational efficiency, in response to the fluctuating energy conditions. We will demonstrate in Section~5 how this method significantly contributes to long-term dependability by optimizing energy use.




\begin{table}[tp]
    \centering
    \caption{State space definitions}
    \vspace{-0.2cm}
    \begin{tabular}{|l|l|c|}
    \hline 
    Symbols & Descriptions & Values\\ \hline 
    $E_{now}$ &  Current battery level  & \multirow{2}{*}{\parbox{2.5cm}{0: depleted, 1: low,\\2: high, 3: full}} \\ \cline{0-1}
    $E_{last}$ &  Last 10 epochs mean energy &  \\ \hline
    $P_{harv}$      & Harvesting power level &  0: low, 1: mid, 2: high \\ \hline
    $L$     & Num of learners executed &  $[0,1,...,N]$ \\ \hline
    $R$     & Inference request & 0: No, 1: Yes \\ \hline 
    \end{tabular}
    \label{tab:state-space}
\end{table}

\begin{equation}
    Reward(s,a) = 
    \begin{cases}
        \Delta_{acc} - \beta (E_{max}-E_{now})  & a=1 \\
        -p_{miss}    &   R=1 \text{ and } a=0    \\
    \end{cases}
    \label{eq:rewards}
\end{equation}


\noindent
\textbf{Concurrent Inference and Training:}
Dataset shifts in real-world scenarios~\cite{lee2020learning} can diminish the accuracy of a pre-trained ensemble once deployed on a microcontroller. Therefore, the capacity to adapt to new data distributions is crucial for edge intelligent applications in delivering reliable predictions. Nevertheless, previous works on on-device training usually consider single-instance models and overlook the issue of dependability: indeed, single-instance models cannot perform any inference task while they are undergoing a retraining stage, thus reducing system dependability. In this work, we address this challenge by leveraging the proposed E-QUARTIC architecture and by extending the functionalities of the presented task scheduler.

\begin{figure}[tp]
    \centering
    \includegraphics[width=\linewidth]{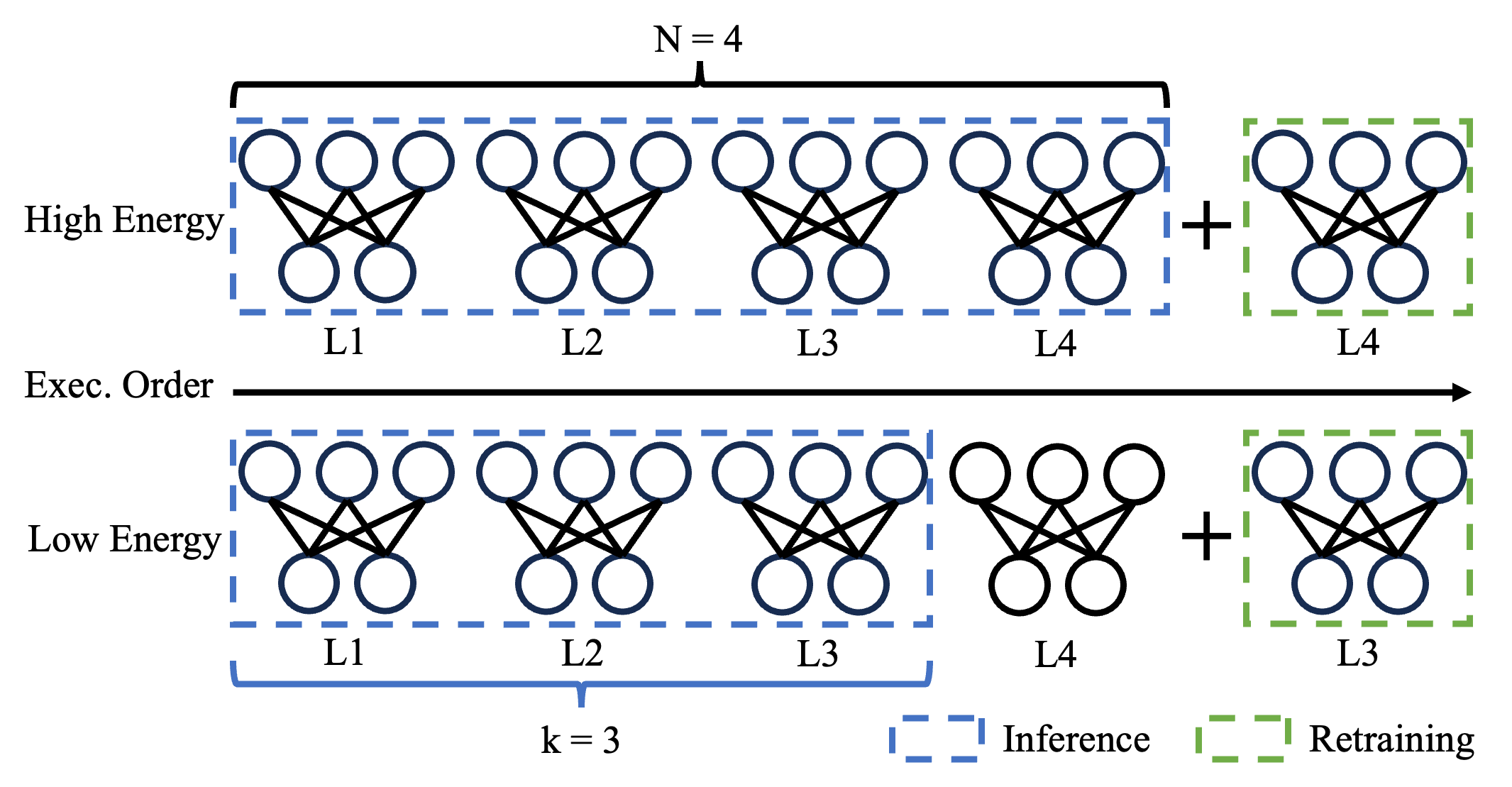}
    \vspace{-0.5cm}
    \caption{Concurrent training and inference stages in low and high energy conditions, for $N=4$ and $k=3$.}
    \vspace{-0.3cm}
    \label{fig:inference-training}
\end{figure}

In particular, the proposed scheduler can accommodate training stages without affecting inference executions both in the case of high and low energy levels. In both cases, we focus our analysis on the retraining opportunity of one CNN instance at a time. We also consider the retraining stage of fully-connected layers only, hence significantly limiting energy overheads. Finally, we observe that the retraining of one weak learner still requires a forward pass, which can also used to produce the E-QUARTIC inference prediction. Consequently, we propose two alternative energy-aware retraining approaches, summarized in Figure~\ref{fig:inference-training}, where blue and green boxes represent learners used for inference and retraining, respectively. In the case of high energy levels, we follow the previously described scheduler policy and execute all the weak learners of E-QUARTIC. One CNN instance can then be retrained for negligible energy overheads (see Section~5). Instead, in the case of low energy levels, only $k<N$ instances are run, followed by one of them undergoing the retraining stage. 
In this second case, we enable concurrent inference and training while prioritizing inference over training. This approach reduces runtime requirements and energy consumption while maintaining inference reliability.

\section{Evaluation}

\subsection{Experimental Setup}

\textbf{Hardware and energy-harvesting simulation:} We conduct experiments using an STM32L552ZE evaluation board~\cite{stm32-evaluation-board}, which is equipped with an ARM Cortex-M33 processor, 256KB of SRAM, and 512KB of flash memory, operating within a voltage range of 1.7V to 3.6V. A 0.47F supercapacitor serves as the energy storage. An indoor solar harvesting dataset~\cite{sigrist2019dataset} is employed for energy simulations. It comprises voltage and current traces from solar panels deployed at six indoor locations for two years. We train the Q-learning scheduling agent offline using 40-minute energy trace clips from April 1 to April 15, 2018 considering an indoor location with possible direct sunlight up to 20k lux illuminance during the daytime~\cite{sigrist2019dataset}. We include experimental results considering energy traces during which the capacitor undergoes a complete cycle of charging and discharging. We select the voltage and current traces from the dataset that are boosted by the energy harvester, ensuring a stable charging voltage of 4.2V. The energy-aware Q-learning-based scheduler that controls E-QUARTIC executions is trained offline and is then deployed in the microcontroller for dynamic energy experiments.
Two metrics are considered to evaluate the E-QUARTIC performance across the different evaluated designs: \emph{Mean accuracy} represents the average accuracy of successful inference executions and \emph{failure rate} measures the ratio of failed executions over the total number of inference events released over a specific time interval. 

\begin{table}[tp]
\caption{Experimental setups for E-QUARTIC and SOTAs}
    \centering
    \begin{tabular}{|c|l|l|}
    \hline     SOTA                                &       Algorithm       &       Descriptions               \\
    \hline
    E$^{2}$CNN~\cite{ponzina2021e2cnn}  &       Voting         &      Ensemble of 4 learners  \\
    \hline
    Harvnet~\cite{jeon2023harvnet}     &       Early-exit      &       CNN with 3 early exits  \\
    \hline
    Adaptive~\cite{park2023energy}     &       Model pool      &       Pool of 3 CNNs  \\
    \hline
    \sys                               &       Boosting       &       Ensemble of 4 learners  \\
    \hline
    \end{tabular}
    \label{tab:sota}
\end{table}

\noindent
\textbf{Models and datasets:} We evaluate E-QUARTIC on multiple tinyML benchmarks including ResNet-8, MobileNetV1, and DSCNN, as implemented in the MLPerf Tiny Benchmark~\cite{banbury2021mlperf} and considering the CIFAR-10~\cite{krizhevsky2012imagenet}, Visual Wake Word (VWW)~\cite{chowdhery2019visual}, and Google Speech Commands (GSC-10)~\cite{warden2018speech} datasets, respectively.
Models are trained and quantized using 8-bit precision in PyTorch and implemented on the microcontroller using Capuchin~\cite{zhang2022demo} tiny model generator. 

\noindent
\textbf{Baselines:} We compare the E-QUARTIC against the baseline single-instance CNN implementations and the following three recent works: E$^{2}$CNN~\cite{ponzina2021e2cnn}, Harvnet~\cite{jeon2023harvnet}, and Adaptive~\cite{park2023energy}. 
Our evaluation focuses on memory requirements, accuracy, and energy consumption.
We present results on E-QUARTIC implementations including four weak learners, although these results extend to other E-QUARTIC sizes. For fair comparisons, we also consider E$^{2}$CNN designs composed of the same number of instances, while we pick the best-performing implementations for Harvnet~\cite{jeon2023harvnet} and Adaptive~\cite{park2023energy} that can meet the memory requirements of the target hardware board. A summary of the considered baselines is included in Table~\ref{tab:sota}.

\begin{table}
\caption{Comparing~\sys~with single-instance and SOTA baselines in terms of memory and computing requirements, and accuracy. \sys~improves memory and accuracy, without incurring any computing overhead.}
\label{tab:models-e2cnn}
\begin{tabularx}{\linewidth}{@{}lcccc@{}}
\hline
Model(Dataset)   &    SOTA   &  Mem(KB) &  MACs(M) & Acc(\%)  
\\ \hline
\multirow{5}{*}{\parbox{2cm}{ResNet-8\cite{banbury2021mlperf} \\ (CIFAR-10\cite{krizhevsky2012imagenet})}}  &    Baseline                            &     312.5    &       46.8 &         84.0          \\
\cline{2-5}
    &    E$^{2}$CNN\cite{ponzina2021e2cnn}  &     310.8  &  52.3 &                    78.2                     \\
\cline{2-5}
    &    HarvNet\cite{jeon2023harvnet}   &     339.4    &  46.8&                  83.5                      \\
\cline{2-5}
    &    Adaptive\cite{park2023energy}          &     430.2    &    46.8&                84.0                      \\
\cline{2-5}
    &    \sys                                &     \textbf{213.3}    &   46.8&                 \textbf{84.2}                     \\
\hline
\hline

\multirow{5}{*}{\parbox{2cm}{MobileNetV1\cite{banbury2021mlperf} \\ (VWW\cite{chowdhery2019visual})}}    &    Baseline                            &     279.4              &      33.9 & 82.0          \\
\cline{2-5}
    &    E$^{2}$CNN\cite{ponzina2021e2cnn}  &     277.2     &      34.9 & 81.4                         \\
\cline{2-5}
    &    HarvNet\cite{jeon2023harvnet}   &    287.7         &    33.9 & 82.0                      \\
\cline{2-5}
    &    Adaptive\cite{park2023energy}          &     502.6      &   33.9 &  82.0                      \\
\cline{2-5}
    &    \sys                                &     \textbf{125.0}          &   33.9 &  \textbf{82.8}                      \\
\hline
\hline

\multirow{5}{*}{\parbox{2cm}{DSCNN\cite{banbury2021mlperf} \\ (GSC-10\cite{warden2018speech})}}    &    Baseline                            &     44.4       &     55.7    &      94.0          \\
\cline{2-5}
    &    E$^{2}$CNN\cite{ponzina2021e2cnn}  &     44.2           & 70.7        &      93.3                        \\
\cline{2-5}
    &    HarvNet\cite{jeon2023harvnet} & 51.7 &     55.7          &  94.0                     \\
\cline{2-5}
    &    Adaptive\cite{park2023energy}          &     87.4     & 55.7  &     94.0                      \\
\cline{2-5}
    &    \sys                                &     \textbf{31.4}        & 54.6  &    \textbf{94.9}                      \\
\hline
\vspace{-0.5cm}

\end{tabularx}
\end{table}


\begin{figure*}
  \includegraphics[width=\textwidth]{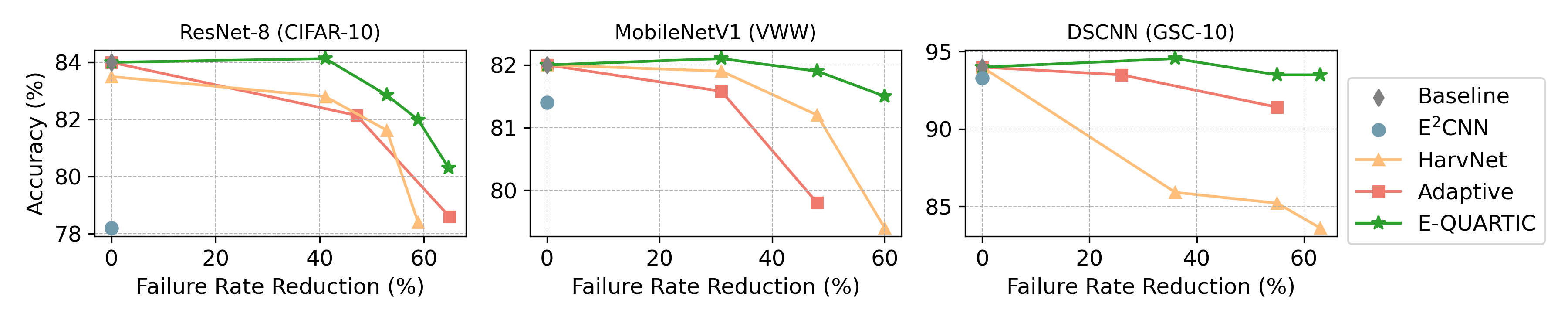}
  \vspace{-0.8cm}
  \caption{Mean accuracy-failure rate reduction Pareto graphs}
  \label{fig:pareto}
\end{figure*}

\subsection{Results}

\textbf{Memory, Accuracy, and Computational complexity:} We compare E-QUARTIC to the considered baselines in terms of memory requirements, accuracy, and computational complexity in \emph{an offline energy-static environment} in~Table~\ref{tab:models-e2cnn}. The baseline methods include single-instance CNN model, E$^2$CNN~\cite{ponzina2021e2cnn}, HarvNet~\cite{jeon2023harvnet}, and Adaptive~\cite{park2023energy}.
Our results demonstrate E-QUARTIC satisfies its offline design goals of (i) high accuracy, (ii) equal or lower memory requirements, and (iii) a similar energy/runtime cost. 
First, we compare E-QUARTIC to the baseline single-instance CNN model and E$^2$CNN~\cite{ponzina2021e2cnn}. E-QUARTIC outperforms these two baselines by 30-55\% memory savings, up to 22\% computational complexity reductions, and up to 6\% accuracy improvements. These improvements are mainly due to the boosting-based ensembling methods employed in E-QUARTIC.  
When compared to SOTA methods HarvNet~\cite{jeon2023harvnet} and Adaptive~\cite{park2023energy}, our results show that E-QUARTIC exhibits similar runtime requirements but it outperforms them by up to 30-75\% memory savings and around 1\% accuracy improvements. The memory savings demonstrate E-QUARTIC's advantages in dealing with memory scaling issues. 
As the granularity of energy management increases, i.e., the ensemble size, the exit number, and the model number increase for the three methods, respectively, E-QUARTIC exhibits similar or even much lower memory requirements. In contrast, HarvNet~\cite{jeon2023harvnet} and Adaptive~\cite{park2023energy} will eventually exhaust their on-device memory due to their method-specific structure overheads, i.e., the early-exit classifiers and the multiple models. For example, the 512KB flash memory has the capacity to store only three distinct models for Adaptive~\cite{park2023energy}, while E-QUARTIC can store up to $N=5$ learners, thus demonstrating stronger memory adaptability as the granularity of energy management increases.
Furthermore, E-QUARTIC memory savings facilitate the storage of the Q-Table required by the scheduler for its dynamic inference executions. The Q-Table's negligible memory overhead of a few tens of bytes will eventually be offset by the huge E-QUARTIC memory savings.

\begin{figure}[tp]
    \centering
    \includegraphics[width=\linewidth]{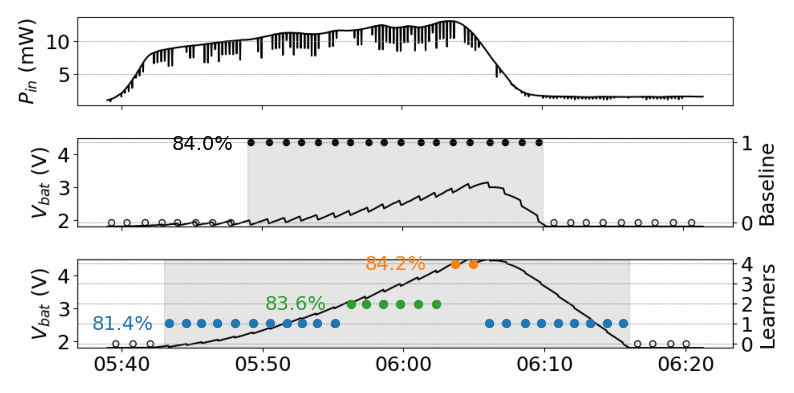}
    \caption{Harvested power trace (top), battery levels and inference traces of ResNet-8~\cite{banbury2021mlperf} (middle) and the corresponding E-QUARTIC implementation (bottom). }
    \label{fig:resnet-inference}
    \vspace{-0.3cm}
\end{figure}

\noindent
\textbf{Mean accuracy/Failure rate trade-off:} We herein evaluate the performance of \sys~in \emph{a dynamic energy environment}.
First, we compare E-QUARTIC with the baseline single-instance CNN by evaluating how much battery lifetime extension E-QUARTIC offers, measuring the accuracy penalty of these improvements. Our findings are illustrated in Figure~\ref{fig:resnet-inference}, with the first plot showing the harvested power levels while the second and third ones depicting the battery level and inference history of the single-instance model (ResNet-8~\cite{he2016resnet}) and E-QUARTIC, respectively.  Solid circles represent successful inference executions and empty ones represent failures. Grey areas indicate the system is active.
The bottom plot indicates how E-QUARTIC can dynamically adapt the energy levels, running ensembles of different sizes. In particular, it achieves 81.4\%, 83.6\%, and 84.2\% accuracy when executing one, two, and four learners, respectively, compared to the baseline 84.0\% accuracy. 
Our results show that E-QUARTIC can react fast to the increasing availability of energy and, by smartly executing ensembles of a variable number of instances, extend the active time. ~\sys~turns on earlier in lower battery voltage at 05:44 before 05:50 and smartly extends battery life from 06:10 to 06:16 by executing fewer learners as the battery level drops. As a result, E-QUARTIC reduces failure rate by over 50\% for just an average 2\% accuracy drop. 

We also conduct similar experiments on different SOTA variant designs and illustrate in Figure~\ref{fig:pareto} the trade-off between average runtime accuracy and failure rate reduction across different ensemble sizes and three models. Points from left to right in each curve correspond to designs with higher numbers of instances (i.e., E-QUARTIC and Pool) or higher numbers of exits (Early-exit). For completeness, we also consider the baseline single-instance CNN model and E$^2$CNN~\cite{ponzina2021e2cnn}, but each of them is presented in a single point in the figures due to their lack of energy-adaptive designs.
We observe that reductions in failure rates are generally associated with accuracy degradation, as the higher granularity in the model design allows them to better adapt to energy variations. Interestingly, this is not the case for E-QUARTIC when composed of two only instances. In fact, such an E-QUARTIC design improves both metrics at the same time: when compared to the baseline single-instance implementation, failure rate is reduced by 30-40\%, with output quality that is still 0.2-0.5\% higher. 
In general, E-QUARTIC dominates all the SOTA methods as it achieves higher failure rate reductions while preserving higher accuracy.
On average, E-QUARTIC achieves accuracy improvements of 1.2-1.7\%, 1.0-1.4\%, and 2.1-8.3\% and with respect to the considered baselines for similar failure rate reductions for three models, respectively. Dually, E-QUARTIC can reduce failure rate of the considered baselines by 30-40\% for similar accuracy levels.

\noindent
\textbf{On-device Training:} Finally, we evaluate the on-device training performance from energy and performance perspectives. We present the results of the two different approaches described in Section~4 (i.e., either run $N$ or $N-1$ weak learners and then retrain one). When executing all of them for inference, we measure energy and performance overheads of 0.04\% per retraining epoch and sample. Notice how this overhead is negligible even for larger batch sizes (e.g., 10/50 samples per batch), and only memory constraints are the real limitations for E-QUARTIC retraining. Instead, when executing E-QUARTIC with $N-1$ instances, concurrent inference and retraining can be achieved at a negligible accuracy cost of 0.4\%, but saving 25\% of runtime and energy cost.

\section{Conclusion}

In this paper, we proposed E-QUARTIC, a novel energy-efficient ensembling method for edge AI. It combines pruning, replication, and boosting methods to build ensembles of CNNs that improves accuracy when compared to single-instance baselines and SOTA solutions while dramatically reducing memory requirements. We leverage the E-QUARTIC architecture to propose a new strategy to run inference in constrained embedded systems to adapt to dynamic energy environments. We demonstrate that E-QUARTIC can dynamically schedule inference tasks based on a lightweight analysis of energy levels achieving up to 40\% battery lifetime extension for negligible accuracy drops.




\section*{Acknowledgements}
This work was supported in part by the National Science Foundation under Grants \#2112665 (TILOS AI Research Institute), \#2003279, \#1911095, \#1826967, \#2100237, \#2112167, and in part by PRISM and CoCoSys, centers in JUMP 2.0, an SRC program sponsored by DARPA.

\bibliographystyle{acm}

\end{document}